\documentclass[10pt]{article}

\usepackage{amsmath}
\usepackage{amssymb}

\usepackage{graphicx}

\usepackage{cite}

\usepackage{color}

\usepackage[labelfont=bf,labelsep=period,justification=raggedright]{caption}

\makeatletter
\renewcommand{\@biblabel}[1]{\quad#1.}
\makeatother

\date{}

\begin{document}

\begin{flushleft}
{\Large
\textbf{Fashion, Cooperation, and Social Interactions}
}
\\
Zhigang Cao$^{1,\ast}$,
Haoyu Gao$^{1}$,
Xinglong Qu$^{1}$,
Mingmin Yang$^{1}$,
Xiaoguang Yang$^{1}$,
\\
\bf{1} Key Laboratory of Management, Decision \& Information Systems, Academy of Mathematics and Systems Science,
        Chinese Academy of Sciences, Beijing, China\\
$\ast$ E-mail: zhigangcao@amss.ac.cn, gaosteveneric@gmail.com, quxinglong10@mails.gucas.ac.cn, yangmingmin2005@126.com, xgyang@iss.ac.cn
\end{flushleft}

\section*{Abstract}
Fashion plays such a crucial rule in the evolution of culture and society that it is regarded as a second nature to the human being. Also,
its impact on economy is quite nontrivial. On what is fashionable, interestingly, there are two viewpoints that are both extremely widespread but almost
 opposite: conformists think that what is popular is fashionable, while rebels believe that being different is the essence. Fashion color is fashionable
 in the first sense, and Lady Gaga in the second. We investigate a model where the population consists of the afore-mentioned two groups of people that
  are located on social networks (a spatial cellular automata network and small-world networks). This model captures two fundamental kinds of social
  interactions  (coordination and anti-coordination) simultaneously, and  also has its own interest to game theory: it is a hybrid model of pure competition
  and pure cooperation. This is true  because when a conformist meets a rebel, they play the zero sum matching pennies game, which is pure competition.
  When two conformists (rebels) meet, they play the (anti-) coordination game, which is pure cooperation. Simulation shows that simple social interactions
  greatly promote cooperation: in most cases  people can reach an extraordinarily high level of  cooperation, through a selfish, myopic, naive, and local
  interacting dynamic (the best response dynamic). We find that degree of synchronization also plays a critical role, but mostly on the negative side.
  Four indices, namely cooperation degree, average satisfaction degree, equilibrium ratio and complete ratio,  are defined and applied to measure people's cooperation levels from various angles.
 Phase transition, as well as emergence of many interesting geographic patterns in the cellular automata network, is also observed.
\section*{Introduction}
Fashion is a very interesting phenomenon that plays a critical role in economy. Firstly, fashion is a huge industry. Despite the complication to give a
satisfactory definition of the fashion industry, its global market size is estimated at some two hundred billion US dollars \cite{o07,e11}. The importance
of the fashion industry can be further justified by the following facts \cite{o07}: fashion industry ``is one of the few industrial segments that have
remained a constant world economy contributor with an annual growth rate of approximately 20 percent. [...] The luxury fashion sector is the fourth largest
revenue generator in France; and one of the most prominent sectors in Italy, Spain, the USA and the emerging markets of China and India. The sector is
currently one of the highest employers in France and Italy." Secondly, fashion serves as a constant and  efficient consumption stimulator.  It is well
known that the purchases of most consumer durables are replacement ones \cite{v10}. For instance, the shoes that we buy are usually not our first ones, the
aim that we buy them is to replace our old ones. However, a not so environmentally friendly fact is that  a significant percentage of replaced products
still function very well \cite{v10}. It is becoming more and more popular that people discard their old possessions not because they fail to meet their
physical needs, but because they are not fashionable any longer. If all people had purchased new goods only when their old ones were completely broken,
then the world economy would have grown much slower.

Not only in economy, but also behind many phenomena in society, education, politics, arts and academics, fashion is a factor that cannot be neglected. A
recent research shows that even charitable donation is a highly subject of fashion \cite{sm08}. Lars Svendsen, a famous philosopher, argues that \cite{s06}
``Fashion has been one of the most influential phenomena in Western civilization since the Renaissance. It has conquered an increasing number of modern
man's fields of activity and has become almost `second nature' to us."  He believes that fashion deserves serious studies from philosophers. We are sure
that as the post industrial society \cite{b73} is coming to reality for more and more countries, the practical functions of commodities and human behaviors
are mattering less and less compared with their social functions. Consequently, fashion is playing an increasingly crucial role.

On what is fashionable, interestingly, there are two almost opposite viewpoints that are both extremely popular.
 One point of view thinks that fashion is  a distinctive or peculiar manner or way. Lady Gaga is regarded as fashionable in this sense.
 The other takes fashion as  a prevailing custom or style. Fashion color is fashionable in this latter sense. This difference has a
  very deep root in psychology, and reveals that people have various desiring or enduring levels of to be how different with the others.
   Following Jackson \cite{j09}, we call the former type of people {\it rebels}, and the latter {\it conformists}.

The phenomenon of fashion has so far attracted some attention from the academical realm, mostly in economics \cite{b92, bhw92, p95, cep07, o07, age12}. However, compared with its great importance, this is far from sufficient.

We shall study fashion through a game-theoretical model, which is called {\it the fashion game}.
  Formally, each fashion game is represented by a triple $\mathcal{I}=(N,E,T)$, where $N=\{1,2,\cdots,n\}$ is the set of agents, $E\subseteq N\times N$ the set of edges, and $T=(\tau_1,\tau_2,\cdots,\tau_n)\in \{C,R\}^N$ the configuration of  types. For each agent $i\in N$, $\tau_i\in \{C,R\}$ is her type: $\tau_i=C$ means that $i$ is a conformist, and $\tau_i=R$ a rebel. For agents $i,j\in N$, they are neighbor to each other if and only if $ij\in E$. If $ij\in E$, then $ji\in E$,  i.e. the network is undirected. $N_i$ is the neighbor set of player $i$, and
$\{0,1\}$ the identical (pure) action set of all players. We use $x_i\in \{0,1\}$ to denote the action of player $i$. Given a pure action profile $X=(x_1,x_2,\cdots,x_n)$, $L_i(X)\subseteq N_i$ is the set of neighboring agents that $i$ likes (w.r.t. $X$), i.e. \begin{equation*}L_i(X)=\left\{\begin{array}{cc}\{j\in N_i: x_j=x_i\} & if ~\tau_i=C\\
\{j\in N_i: x_j\neq x_i\}& if ~\tau_i=R\end{array}\right..\end{equation*}

Similarly, $H_i(X)\subseteq N_i$ is the set of neighboring agents that $i$ hates, i.e. $H_i(X)=N_i\setminus L_i(X)$.
Using $|\cdot|$ to denote the cardinality of a set, the utility function of player $i$ can be defined naturally as follows:
\begin{equation*}u_i(X)=|L_i(X)|-|H_i(X)|.\end{equation*}

 Utilities of mixed action profiles can be extended as usual. If the utility of an agent is nonnegative, we say that she is {\it satisfied}. An action profile that all agents are satisfied corresponds clearly to a Nash equilibrium.

 As far as we know, the above fashion game is proposed by  Young (\cite{y01}, 2001, p.38) and Jackson (\cite{j09}, 2008, p.271), independently. Obviously, this game is an extension of the famous matching pennies game, which is stated below.
 \begin{center}\begin{tabular}{|c|c|c|}
\hline
&H&T\\
\hline
H&$(1,-1)$&$(-1,1)$\\
\hline
T&$(-1,1)$&$(1,-1)$\\
\hline\end{tabular}.\end{center}

  In fact, in matching pennies, the row player is a conformist, and the column player a rebel. A special case of the fashion game, a dyad with one conformist and one rebel, is exactly the matching pennies game. Just like matching pennies, the fashion game always has a mixed Nash equilibrium: all agents play half 0 and half 1. The existence of pure Nash equilibrium, however, cannot be guaranteed, and it is NP-hard to check whether a fashion game on a general network has a pure Nash equilibrium or not \cite{cy11}. Consequently, it is impossible to compute a pure Nash equilibrium efficiently when it does exist, unless P=NP (i.e. the set of problems admitting deterministic polynomial time algorithms equals that of problems admitting non-deterministic polynomial time algorithms, a statement widely conjectured as impossible to be true. Extensive discussion of this conjecture can be found in any textbook about computational complexity).

 It is valuable to note that the fashion game, though very simple, is a typical heterogeneous model. There are two types of players, and actually three {\it base games} played:  When a conformist faces a conformist, they play the pure coordination game; When two rebels meet, they play the pure anti-coordination game; And when a conformist confronts a rebel, they are in the exact game of matching pennies.

 It is widely accepted that competition and cooperation are the two eternal topics in game theory.  Zero sum games (or more generally, constant sum games)
 are polar examples for competition. This is why they are also called {\it strictly competitive games} (cf. \cite{or94}, p.21). Common interest games (a.k.a. team games) are polar examples for cooperation, where the preference rankings of pure action profiles for all players are the same. Among the three base games of the fashion game, the pure coordination game and the pure anti-coordination game are both common interest games,  while matching pennies is a zero sum game.  For general normal form games, competition and cooperation are both embodied. This is a kind of {\it vertical hybrid}, and is the key observation of \cite{kk09}.  The fashion game, on the other hand, is a kind of {\it horizontal hybrid} of competition and cooperation. This feature determines that the fashion game has a very special interest to game theory.

The fashion game can be safely classified into {\it network games}, a typical multi-disciplinary field that rests at the intersection of social economics,
 social physics, theoretical biology, and algorithmic game theory \cite{j09,jz14,gg10,sf07,kls01}. Studying closely related models, it is a pity that
 researchers from different fields seldom cite each other so far.
Although the exact model of the fashion game does not draw much attention today, there has been a lot of related work. In social economics, the
coordination game, one side of the fashion game,  has been  extensively studied \cite{bs96,e93,b93,b95,m00,y01,bd01,jw02,w02,gv05}. The other side of the
fashion game, the anti-coordination game, however, attracts very little attention \cite{b07, blgv04, l09, cy12}. In the field of social physics, for models
where agents are homogenously conformists or homogenously rebels, there has been a large number of references, which are impossible to fully survey here.
They are called ``majority game" (cf. \cite{ai93,c09}) and  ``minority game" (cf. \cite{cz97,cmz04}), respectively. It is well known that the minority game
has been used to study the financial and stock market since very soon after its birth. Marsili \cite{m01} noticed
 early enough that people do not necessarily play the minority game in stock market, because except for a few  ``market fundamentalists", most people are ``trend followers".
 And thus instead, the minority-majority game  is much more appropriate. Following Marsili, there are a dozen of papers \cite{mgm03,g06a,g06b,g06c,c08,c09,csw08,cggg08,lsd10}.
 Another extensively studied model that is closely related with the anti-coordination game is the snowdrift game (a.k.a. the hawk-dove game, see the excellent survey \cite{sf07}).
 Related work in the fields of social learning and opinion dynamics includes \cite{ao10,gj10,cyqy11,g04,g08}. In statistical physics, another similar model
 is the generalization of the Kuramoto model with conformist  oscillators and contrarian  oscillators (\cite{hs11a,hs11b,hs12}).
It is valuable to remark that, in the afore mentioned papers, neither the relation of their models with  the phenomenon of fashion nor the relation with
the matching pennies game is noticed.
 For recent work on repeated matching pennies from
the perspective of behavioral science, please refer \cite{er11}.
 A brand new perspective on matching pennies is that it can be interpreted as a (symmetric) Predator-Prey game \cite{sb08, js12}. A quite complete literature review of ``Social Influence, Binary Decisions and Collective Dynamics", a classical topic in sociology and also closely related with this paper, can be found in \cite{lw08}.

The main concern of this paper is, in a world where each agent is fashionable (in one of the  possible two ways), selfish (cares only about the welfare of her own), extremely naive and myopic (has only one step memory and does not look forward), and has very limited information (that of herself and her neighbors), to what extent can cooperation be reached through social interaction? Will things be even worse?

Our finding is, in general, quite encouraging: an extraordinarily high level of cooperation can be reached through a simple updating rule, the {\it best
response dynamic}. That is, social interactions generally promote cooperation (compared with the initial settings that are uniformly random, i.e. each
agent takes action 1 with probability 0.5 and action 0 with probability 0.5). In very rare cases, the discouraging result that social interactions prohibit
cooperation can also be observed. Degree of synchronization, captured by updating probability in the best response dynamic
 that will be introduced in the next section, also plays a critical role, but mostly on the negative side. We remark that the negative effect
 of synchronization is observed in \cite{lw08} too, for the anti-coordination game on a complete network.

\section*{Methods}
Most of our conclusions are derived through a two dimensional (stochastic) cellular automaton. This spatial structure is simple enough and can serve very well for a first step study. After discussion of this particular case, more extensive analyses are done for the more general and more realistic small-world networks \cite{ws98}, where most of  our conclusions are confirmed. Since it is notoriously hard to rigorously analyze a two dimensional cellular automaton (even a deterministic one), and the negative result of \cite{cy11} tells us that there is no good characterization of the fashion game, simulation is our natural choice.

Three things need to be clearly stated: the underlying network, the initial settings, and the updating rule. In this section, we shall only introduce  the
particular cellular automata network. Small-world networks and the corresponding simulation settings will be introduced in a later section. The updating
rule, as well as the four indices that will be introduced soon, is universal.
\subsection*{Primary simulation settings}
The cellular automata network is a special 8-degreed regular graph. It is a sheet of grids, where each grid stands for an agent. For each agent, the eight grids that are touching it are her neighbors (that is, we are taking the Moore neighborhood). The sheet is finite but unbounded: each leftmost agent has three neighbors on the rightmost, and each rightmost agent has three neighbors on the leftmost. Likewise, each uppermost agent has three neighbors on the lowermost, and each lowermost agent has three neighbors on the uppermost. Intuitively, we can imagine this world as  a torus. In the next section, all the simulations (except for the ones in Fig. 6)  are done on such a torus of size $41\times 41=1681$.

To characterize the initial settings and the updating rule, we need two parameters: the rebel ratio $r$ and the updating probability $p$. They are the only parameters for simulations on the cellular automata network.

For each initialization, each agent has a probability of $r$ to be a rebel, and $1-r$ to be a conformist. This is done for all agents independently. $r=0$ is the case of all conformists, and $r=1$ that of all rebels. Since the network is quite large, $r$ can be roughly taken as the percentage of rebels. It is found that the percentage of rebels matters a lot to the fashion game. For a special instance, the all-conformist case is completely different from the all-rebel one.

Actions are initialized uniformly, that is,  each agent takes an action of 1 and 0 equally likely (with probability 0.5). Time elapses discretely. At each time step, each agent checks if she is satisfied with the previous action profile. If so, her action keeps unchanged. Otherwise, she switches to the other action with a probability $p$.
Intuitively, $p$ measures the degree of synchronization. $p=1$ is exactly the synchronous best response dynamic. When $p$ is infinitely small, we know that at each step there is at most one agent switching her action. To be precise, the probability that two or more agents switch actions is second-order infinitely small (w.r.t. $p$). This can be roughly taken as the asynchronous best response dynamic.
The introduction of parameter $p$ allows us to compare the cases of synchronous updating, asynchronous updating, and all the middle cases. It turns out that $p$ also matters a lot, usually on the negative side.

\subsection*{Four indices}
We use {\it cooperation degree} as a main index to measure the level of cooperation between agents. Formally, the cooperation degree for any configuration of actions in a fashion game is defined as the percentage of satisfied agents. Clearly, Nash equilibrium corresponds to an action configuration with cooperation degree of 1. This index can be roughly taken as an approximation of Nash equilibrium. Since Nash equilibrium is not guaranteed, this choice is quite natural. Three other indices are also used, namely  {\it  average  satisfaction degree}, {\it equilibrium ratio}, and {\it complete ratio}.

The satisfaction degree of any agent is the percentage of neighbors that she likes. Obviously, satisfaction degree is an extension of the concept of satisfaction: an agent is satisfied if and only if her satisfaction is no less than 0.5. The average satisfaction degree for an action configuration is simply defined as the average of all agents' satisfaction degrees.

The equilibrium ratio of a group of simulations is defined as the percentage of simulations that pure Nash equilibrium is reached. Recall that pure Nash equilibrium is not guaranteed, and even if it exists, it may not be reached by the best response dynamic. The equilibrium ratio also measures the degree of cooperation, because Nash equilibrium can be treated as stable cooperation. The other advantage of this index is that when we investigate phase transition, it is much sharper than cooperation degree and average satisfaction degree.

Given an action configuration, an agent is said to be completely satisfied if and only if her satisfaction degree is 1. The complete ratio of an action configuration is defined as the percentage of completely satisfied agents.

\subsection*{Simulation size}
All results (except for Fig. 4 and Fig. 5) in the next section are based on 420 groups of simulations: the rebel ratio $r$ takes 21 values, 0, 0.05, 0.1, $\cdots$, 1, and the updating ratio $p$ takes 20 values,  0.05, 0.1, $\cdots$, 1. For each of the 420 combination of parameters, 10 simulations are done. For each simulation, it is stopped at the 500-th step. For each combination of parameters, we display the average of the corresponding 10 final values.

It may seem at the first sight that stopping at the 500-th step and taking an average of only 10 simulations are not sufficient for such a large network. However, it turns out that this is enough, because the dynamic is surprisingly both fast and robust.

\section*{Main Results}

\subsection*{Cooperation degree}

First of all, let's calculate the expected cooperation degree of the initial configuration. This value is equivalent to the probability that an agent is initially satisfied. Since each agent has 8 neighbors in total, and she is satisfied if and only if the number of agents who take the same action as she does  is at least 4 when she is a conformist, and at most 4 when she is a rebel, and each agent takes an action equally likely from $\{0,1\}$, we know that this probability is $(1/2)^8(C^8_8+C_8^7+C_8^6+C_8^5+C_8^4)=(1/2)^8(C^0_8+C_8^1+C_8^2+C_8^3+C_8^4)=163/256\doteq 0.64$.

The average of the 420 cooperation degrees in our simulations is calculated to be 0.97.
The 420 values are displayed in the color map of Fig. 1, where different colors represent different values, as shown in the right bar.

\begin{figure}[!htb]
  \centering
  \includegraphics[width=10cm]{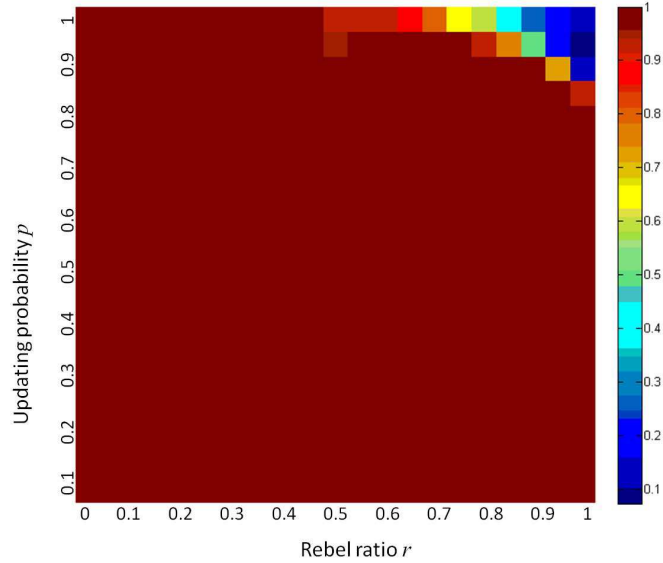}
  \caption{{\bf Cooperation degree.} The fashion game can reach high degree of cooperation through best response dynamics.}
  \label{fig1}
\end{figure}

The most prominent observations from Fig. 1 are as follows.

(1) For most parameter combinations, stable cooperation, i.e. a pure Nash equilibrium, can be reached. This tells us that although pure Nash equilibrium cannot be guaranteed generally, it does exist for most cases in the cellular automaton world, and it can be reached by simple adaptive dynamics: best response dynamics. We remind the reader that, for general games, even if pure Nash equilibrium exists, there is no general simple adaptive dynamic that always leads to one \cite{hm03}. The only theoretical result we know is that best response dynamic (in fact, better response dynamic) always leads a potential game to a pure Nash equilibrium \cite{ms96}. However, the fashion game is not a potential game, even in the cellular  automaton structure. This is because pure Nash equilibrium is guaranteed in potential games, which is not true in the fashion game.

It is also valuable to remark that for the synchronous best response dynamic (i.e. $p=1$), it may not lead the fashion game to a pure Nash equilibrium, even if it exists. In fact, this always occurs when all agents are rebels (a case which is a potential game and thus guarantees the existence of a pure Nash equilibrium, cf. \cite{b07}) and they take the same action initially. It is obvious that at each step, no agent is satisfied, and thus all agents switch to the other action simultaneously, leading to the other state where no agent is satisfied either. The configuration will oscillate between the two extreme states, and no agent is ever satisfied at all. This illustrates the most terrible situation that no one wants to see. Synchronization, of course, plays a critical role. In fact, asynchronous best response dynamic (i.e. $p$ is infinitely small), as shown in Fig. 1,  leads to a pure Nash equilibrium almost surely, and for any $p<1$, the corresponding best response dynamic leads to a pure Nash equilibrium with a positive probability. This will be explored more extensively in later subsections.

(2) Taking into account the fact that the network size is 1681 and in most cases the best response dynamics converge within 500 steps, the convergence is remarkably fast.

(3) Bad cooperation occurs only when the percentage of rebels $r$ and the updating probability $p$ are both high. To investigate this more clearly, we put the upper-right corner of Fig. 1 in another way, as shown in Fig. 2.

\begin{figure}[!htb]
  \centering
  \includegraphics[width=8.3cm]{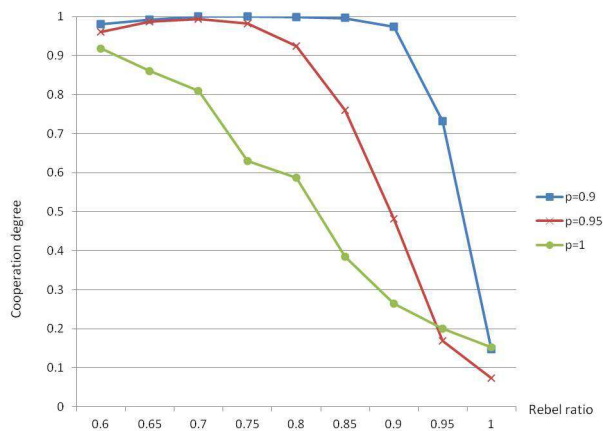}
  \caption{{\bf Bad cooperation.} Cooperation degree is a decreasing function of rebel ratio $r$ when $p$ is large.}
  \label{fig2}
\end{figure}

It can be observed that the cooperation degree is always a decreasing function of the rebel ratio, which means that when the degree of synchronization is high, rebels, in general, are a block for cooperation. This is consistent with the theoretical results that when all agents are conformists, stable cooperation is always possible (i.e. pure Nash equilibrium exists), but when a portion of rebels are added, this may not be true.
It can also be observed that higher probability of updating in general (but not always) means worse cooperation. This is also consistent with our argument in (1), i.e. asynchronization  is better than synchronization. These observations will be further justified by more extensive simulations of the next subsection.

(4) For some combinations of parameters (upper-right corner of Fig. 1), the final cooperation degrees can be  even worse than the initial ones. This tells us that best response dynamics with high updating probabilities may not be always good for the whole system, even if updating itself is costless (i.e. agents don't pay for switching actions).

\subsection*{Average satisfaction degree}
 Since the initializations of all agents are independent, it is obvious that the initial expected average satisfaction degree is $0.5$.
 Fig. 3 is also based on the same 4200 simulations, where each simulation is terminated at the 500-th step. However,  only $11\times 21=$231 values are displayed, each as an average satisfaction degree of 10 corresponding simulations.

\begin{figure}[!htb]
  \centering
  \includegraphics[width=8.3cm]{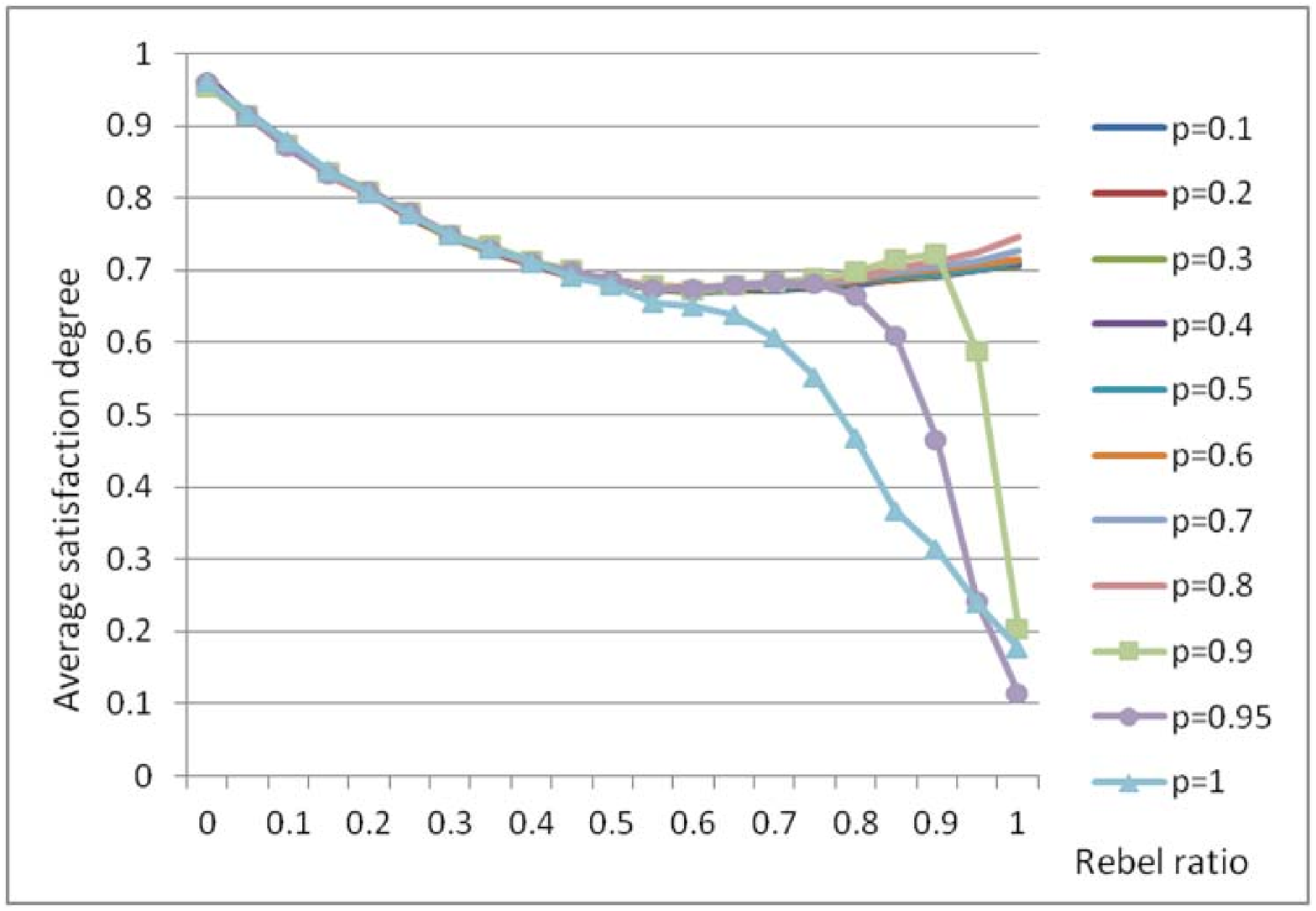}
  \caption{{\bf Average satisfaction degree as a function of rebel ratio.}}
  \label{fig3}
\end{figure}

Below is the analysis of this figure.

(1) In most cases, the average satisfaction degree is rather promising. In fact, the overall average satisfaction degree (i.e. average of the 231 values) is 0.73, much higher than the initial expected value of 0.5. This is consistent with our observations in the last subsection, and justifies our claim once again that, in general, various best response dynamics can promote cooperation.

(2) The overall average satisfaction degree, 0.73, is not that high compared with the overall cooperation degree, 0.97. This is not surprising, because even
 if the cooperation degree is 1, it is very unlikely for the corresponding average satisfaction degree to be 1, in which case it must be that all agents are
  completely satisfied.

If the above perfect situation occurs, we say that {\it perfect cooperation} is obtained. Perfect cooperation can only possibly be obtained in one extreme case: the all-conformist case ($r=0$).  This deepens our  observation in the last subsection in a way that in most cases, almost all agents are satisfied, not completely satisfied, but satisfied to some degree. We remark that  in the all-rebel case ($r=1$), the configuration where every agent gets complete satisfaction in general does not exist, although pure Nash equilibrium is guaranteed. To be precise, in the all-rebel case, the configuration where every agent is completely satisfied exists if and only if the graph is bipartite. Unfortunately, the two dimensional cellular automaton with Moor neighborhood is not bipartite (there are cycles of length three). However, the two dimensional cellular automaton with von Neumann neighborhood, i.e. each agent has exactly four neighbors, is indeed bipartite.

(3) In the all-conformist case, perfect cooperation is not attainable through best response dynamics. In this case, there are two focal equilibria, i.e. all
 agents taking action 1 and all agents taking action 0, and they are both perfectly good. However, it is very unlikely for simple dynamics to lead the system
 to them. It is imaginable that perfect cooperation is hugely hard, and the main obstacle comes from the underlying network. When the network is a complete
 graph, it may be not that hard to reach perfect cooperation.

(4) Fortunately, in the very case that all agents are conformists, the real equilibria reached through best response dynamics are good enough. The average
satisfaction degree is shown to be 0.96. In fact, we will show later that the equilibrium states when all agents are conformists are usually composed of
impressively large``continents", and the agents that are not completely satisfied all sit on the coastlines.

(5) In most cases, the average satisfaction degree is not sensitive to the updating probability $p$, unless $p$ and $r$ are both large ($p>0.8,r>0.5$). In fact, the eight curves for $r=0.1,\cdots, 0.8$ are almost identical. This is consistent with the observations of the last subsection. However, cooperation degrees are monotonic there, always approximately 1. Here, the average satisfaction degree, as a function of rebel ratio, is richer and more interesting. In this sense, the fact that (in most cases) the average satisfaction degree is insensitive to updating probability is much more striking than that (in most cases) the cooperation degree is insensitive to updating probability.

(6) In most cases ($p<0.8$), average satisfaction degree is a convex function of the rebel ratio $r$, and reaches its lowest value 0.68 at $0.5$.
This says that the case where there are equally number of conformists and rebels is the most difficult situation to cooperate.

This is intuitively reasonable, because an edge between a conformist and a rebel always contributes 0 to exactly one of the agents. The number of these
inter-type edges is expected to reach its maximum when $r=0.5$. In fact, the very special case where every agent has four conformist neighbors and four
rebel ones has exactly an average satisfaction degree of 0.5.  Having this in mind, and considering that the expected initial average satisfaction degree
is also 0.5, the final value of 0.68 is really not disappointing at all. The property that average satisfaction degree is decreasing when $r<0.5$ and
increasing when $r>0.5$ is easy to understand. The convexity of this function, however, is not that intuitive.  As to why the satisfaction degree of the
all-rebel case, 0.71, is much lower than 0.96, the satisfaction degree of the all-conformist case, remember the remark we give in (2). The particular value
of 0.71 will be revisited later.

(7) For large $p$ and large $r$, the average satisfaction degree can be rather low, and may be even much lower than the expected initial value, 0.5.
This is consistent with our observations in the last subsection.

(8) When $r>0.5$, the average satisfaction degree is an increasing function of $r$ for small $p$, and a decreasing function of $r$ for large $p$. This reflects that there may be some phase transition for $p$, which will be explored more extensively in the next subsection.

\subsection*{Phase transition}
To explore the possible phase transition, we did more careful simulations. Our main worries are that (i) 500 steps may not be enough for the cases where $p$ and $r$ are both large to reach a relatively stable state, and (ii) 0.05, the difference between any two adjacent $p$s we take, is too rough for observing phase transition. Based on the above two worries, we change the simulation settings to allow each simulation run 5000 steps, and $p$s are taken more densely. To be precise, we investigate 41 $p$s: 0.8, 0.805, $\cdots$, 1, and 5 $r$s: $0.5,0.6,\cdots, 1$. For each of the $41\times 5=205$ parameter combinations, we take 10 simulations and calculate the average cooperation degree. The results are displayed in Fig. 4.
\begin{figure}[!htb]
  \centering
  \includegraphics[width=8.3cm]{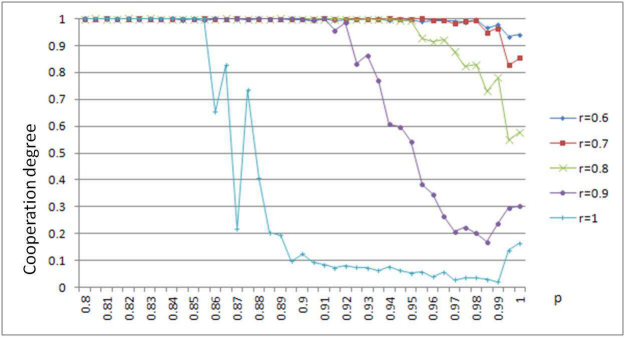}
  \caption{{\bf Phase transition of cooperation degree in p.}}
  \label{fig4}
\end{figure}

\begin{figure}[h]
  \centering
  \includegraphics[width=8.3cm]{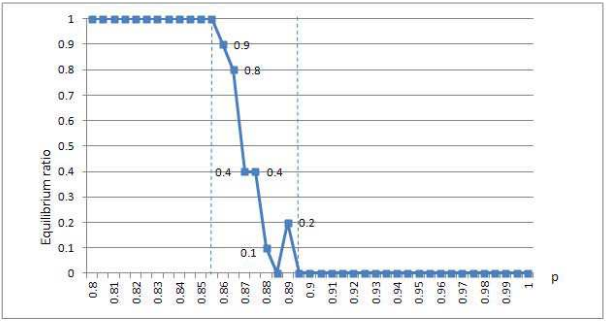}
  \caption{{\bf Phase transition of equilibrium ratio in p.} r=1.}
  \label{fig5}
\end{figure}

 Fig. 4 shows that phase transition does exist, especially for large $r$s: cooperation degree drops from 1 to a very low  level within a small change of $p$. We put the all-rebel case in another way as in Fig. 5, where the index of cooperation degree is replaced by the equilibrium ratio.

Fig. 5 shows that when all agents are rebels, Nash equilibrium can always be reached through best response dynamics if the updating probability $p<0.85$. However, once $p$ exceeds $0.9$, it can never be reached. Within a 5 percent fluctuation  of $p$, the equilibrium ratio drops down sharply from 1 to 0. This is really striking, and we believe that it can be treated as a significant phase transition. The all-rebel case will be explored more deeply in the next subsection, where we shall see that regular patterns emerge steadily regardless of whether Nash equilibrium can be reached or not.

\subsection*{Pattern emergence}
 We show first the all-conformist case and then the all-rebel case. In the figures below, each agent is represented by a small square (for
conformists, it is simply a square; for rebels, there is a dark triangle in this square), red means that the corresponding agent chooses an action of 1 and
green of 0.
\begin{figure}[!htb]
 \centering
  \includegraphics[width=15cm]{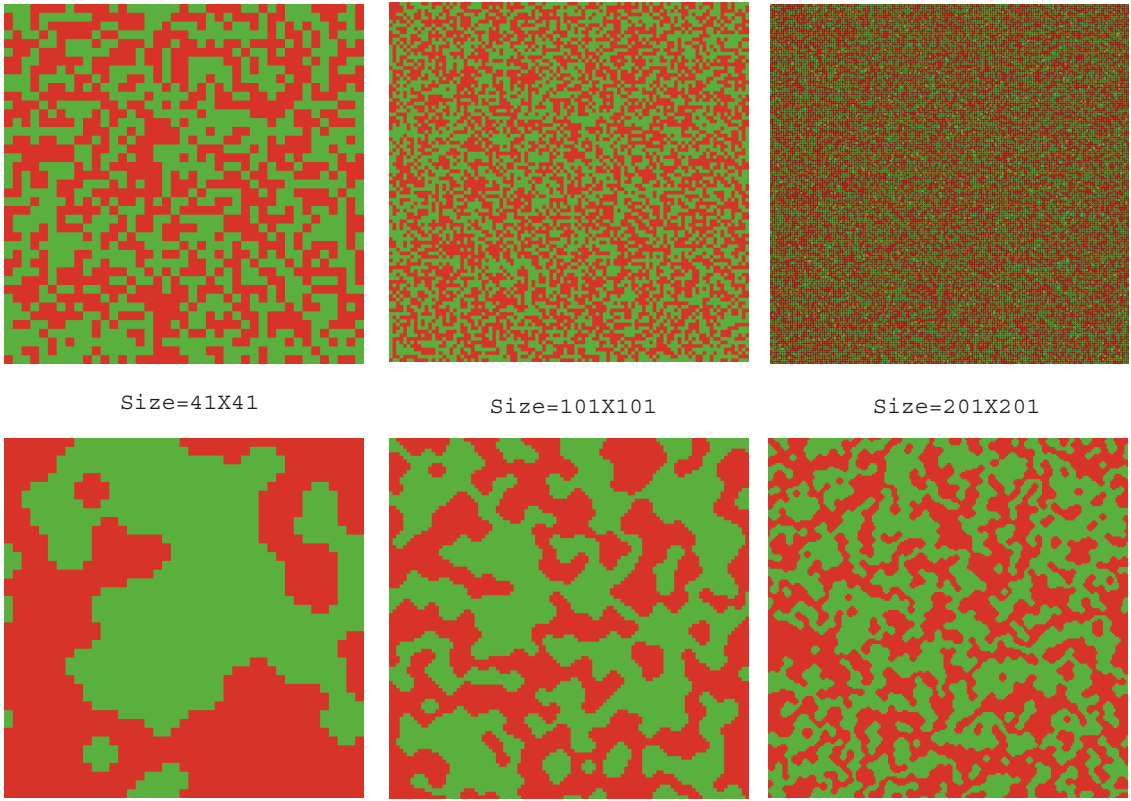}
  \caption{{\bf Emergence of continents.} r=0, p=0.5. The upper three are initial configurations, and the lower three are the corresponding stable ones.}
  \label{fig6}
\end{figure}

We call the set of agents that are connected and share the same action a ``continent". A torus with any configuration can be decomposed into continents.  Fig. 6 shows that toruses with initial configurations have only relatively small continents. However, continents in the stable configurations arrived through best response dynamics are in general impressively large. When the torus is small, the largest continent can even cover more than a half of it. Obviously, an agent is completely satisfied if and only if it is  in the inner part of some continent, and  incompletely satisfied, or unsatisfied, if and only if it is on the ``coastal line", i.e. the boundary of some continent. It can be proved trivially that complete ratio, i.e. the percentage of completely satisfied agents, for any initialization is expectedly $(0.5)^8\doteq 0.0039$. Large continents in the stable configurations imply that the final complete ratio is high. In fact, calculation shows that this value in our simulations (r=0) is in general larger than 0.5. This tells us that if we consider the index of complete ratio as the coordinating ability of best response dynamics, the result is still rather optimistic. ``Lakes", i.e. small continents resting in bigger ones, can also be observed.

\begin{figure}[!htb]
 \centering
  \includegraphics[width=15cm]{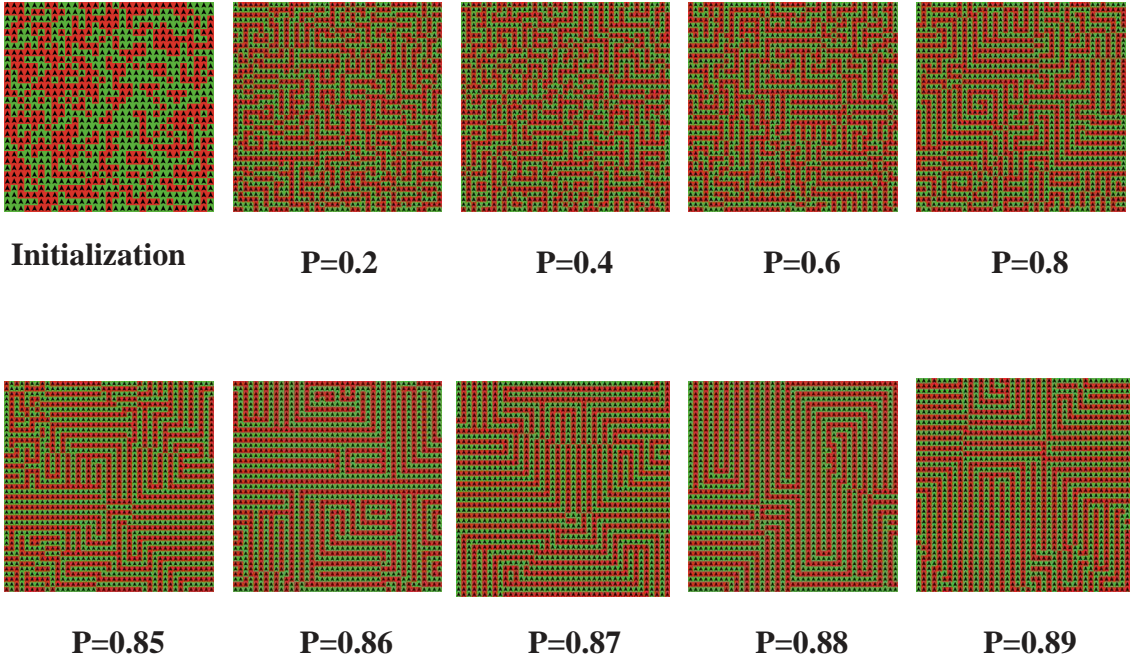}
  \caption{{\bf Emergence of mazes.} r=1,size=41$\times$41.}
  \label{fig7}
\end{figure}

Next, let's turn our attention to the all-rebel case, i.e. $r=1$. An amazing property of this case is that ``mazes" frequently emerge. This is shown clearly in Fig. 7. If an agent is on the ``street" of a maze, rather than on the corner, then she has two neighbors taking the same action as she does, and six ones taking the opposite action. Thus, her satisfaction degree is $6/8=0.75$. Since almost all agents are on streets, and only a few ones on corners, this explains the overall satisfaction degree 0.71 for $r=1$ and $p\leq 0.8$ in Fig. 3 very well.

It should also be noted that every agent in the maze is satisfied, and thus a maze corresponds to a Nash equilibrium. When $r=1$, mazes emerge definitely for $p\leq 0.85$, and never occur for $p>0.9$. This is also implied in Fig. 5. For updating probabilities between 0.85 and 0.9, mazes occur with some positive probability strictly less than 1. It can also be observed that the lengths of streets in mazes generally increase as $p$ increases.

The last part of this subsection is devoted to the emergence for the cases that $r=1$ and Nash equilibrium is not attainable through best response dynamics, i.e. $p>0.9$.

\begin{figure}[!ht]
\begin{center}
\includegraphics[width=15cm]{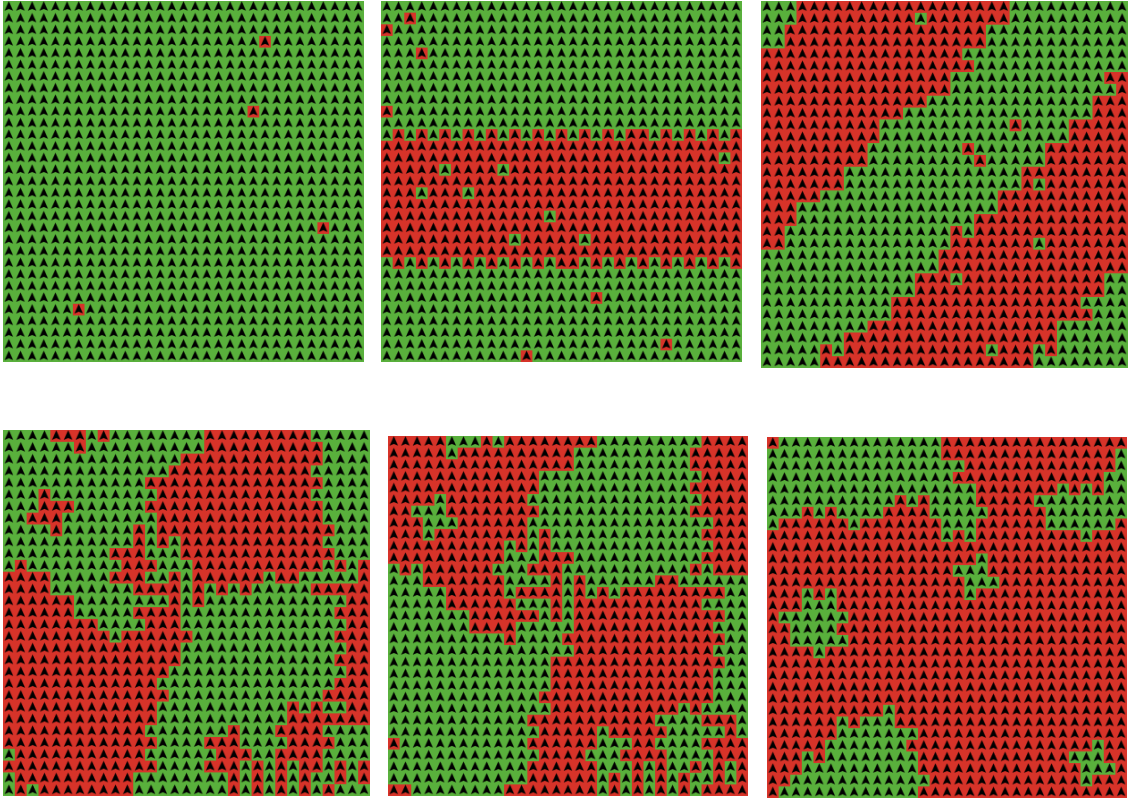}
\end{center}
\caption{
{\bf Emergence of strips, continents, and wheels.}  r=1. p=0.98 for the upper three pictures, and p=1 for the lower three.
}
\label{Figure_8}
\end{figure}

As shown in Fig. 8, regular patterns such as strips (straight ones as well as declining ones) and continents can be frequently observed. Since rebels don't like her neighbors to take the same action as she does, this is not a good thing. And this explains also why the cases that rebel ratios and updating probabilities are both high behave so badly in Fig. 1-5. An extremely terrible situation is the one showed at the upper left corner of  Fig. 8, where almost all agents, except for very few ones (this number is expectedly equal to $(1-p)\times s$, where $s$ is the size of the torus), are completely unsatisfied. The patterns we display in Fig. 8 are the only ones. We note that small wheels can also be frequently observed when $r=1$ and $p=1$.

It you check really carefully the lower three pictures in Fig. 8, you will find that there are only two types of rebels, i.e. the ones that keep switching their actions and the ones that never switch.  This means that the lengths of limit cycles of synchronous best response dynamics in our model are always 2. To put it another way,  synchronous best response dynamics will eventually lead the system to oscillate between two states. The lower left state and the lower middle one in Fig. 8, in fact, compose such a pair.  This is  consistent with the theoretical observation of Cannings \cite{c09} that synchronous best response dynamic always lead a system with all rebels to a limit cycle of length 1 or 2, regardless of the structure. Limit cycles of length 1, i.e. Nash equilibria, however, are never observed by us. This does not mean that they do not exist at all, but that they are really rare. Initial configurations that will eventually reach a Nash equilibrium, can be easily designed. For the simplest instance, we can let the initial configuration be an equilibrium.

Up to now, we have discussed the emergence of regular patterns in the two extreme cases either with all rebels or with all conformists. For the cases in
between, similar patterns can be observed too. In a word, there exists a spectrum and the patterns change over $r$ continuously. When  the rebel ratio $r$
is close to 0, the patterns are more like those in the all-conformist case, and as $r$ tends to 1, the patterns are more similar to those in the all-rebel
case.  But when $r$ is near 0.5, the patterns are quite blurred (we can also say that in these cases there is no pattern at all). What's definite is that
when $r<0.2$ or $r>0.8$, the patterns are rather clear and steady.

All the results of this subsection are obtained with the assistance of the excellent multi-agent programmable software NetLogo \cite{w99}, and our program
can be mailed at request.
\section*{Results on Small-World Networks}
To check that whether the results discussed in the last section, which are derived from a rather special network, are still valid in more realistic networks,
 we did more simulations on small-world networks proposed by Watts and Strogatz (\cite{ws98}). The conclusion is quite promising: to a great extent, our results are rather robust, they are still valid for the more realistic small-world networks. Needless to say, we are unable to discuss the pattern emergence results, because they rely on geographic locations that do not exist in small-world networks. And not surprisingly, we are not able to give convincing explanations to several results as to why they are like that. For instance, we cannot give a similar explanation as to why the average satisfaction degree is 0.71 for the all-rebel case.

Small-world networks are widely accepted as  excellent mimics of the real social networks.  The idea of Watts and Strogatz's algorithm  to generate an
arbitrary small-world network is to derive it from a regular network through ``randomly rewiring" some edges. Roughly speaking, there is a uniform rewiring
probability, which is denoted as $q$ in our paper ($p$ in the paper of Watts and Strogatz), for each of the original edge to be severed and replaced by a
random edge (however, one endpoint of the new edge must  be taken from the two old ones). The advantage of introducing $q$ is that it captures the degree
of randomness of the corresponding network: Larger $q$ implies more randomness. When $q=0$, the network is regular, as in the last section of our paper,
and when $q=1$, the network is completely random (an approximation of the classical ER random network). This gives us a new angle to study the fashion
game, i.e. how randomness affects the cooperation level of agents.

While there are only two parameters in the  simulations of the previous section, namely rebel ratio $r$ and updating probability $p$, there are four in this section. In addition to the two old  ones, there are two new ones, network density (i.e. average degree) $k$ and rewiring probability $q$.  In our simulations,  the network size is fixed at 200, and (1) the network density $k$ takes six values: 8, 18, 28, 38, 48 and 58, (2) the rewiring probability $q$ takes 11 values evenly from 0 to 1, (3) the rebel ratio $r$ takes 11 values evenly from 0 to 1, (4) the updating probability $p$ takes 10 values evenly from 0.1 to 1. For each combination of parameters, the result is based on the average of 10 simulations, and each simulation is terminated at the 500-th step, as set in the previous section.

Calculation shows that the overall cooperation degree is 0.80. This value, though much lower than the corresponding one (0.97) for the cellular automata
network, is still promisingly high, confirming our main result that, in fairly general and realistic scenarios, agents in the fashion game can reach high
level of cooperation through the simple best response dynamic.

\begin{figure}[!ht]
\begin{center}
\includegraphics[width=8.3cm]{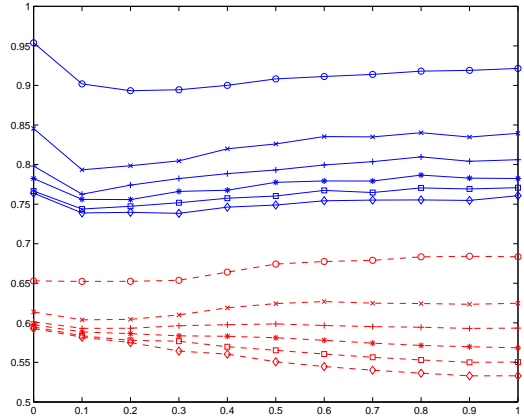}
\end{center}
\caption{
{\bf Effects of rewiring probability $q$ and network density $k$ on cooperation degree and average satisfaction degree.}
 The horizontal axis is for the rewiring probability, solid blue lines for cooperation degree and dashed red ones for average satisfaction degree.
 Different densities of $k=8,18,28,38,48,58$ are represented by $\circ$, $\times$, $+$, $\ast$, $\square$, and $\lozenge$, respectively.}
\label{Figure_9}
\end{figure}

It can also be observed from Fig. 9 that, for all the six cases with different network densities, rewiring probability $q$ always plays a negative role at small values (less than 0.1). But once exceeding the threshold of 0.1, increasing $q$ will be good to the cooperation degree, though the increasing speed of the right part is much lower than the decreasing speed of the left part. The maximum values of cooperation degrees are always obtained at $q=0$, and the values at $q=1$ are only a little bit smaller than the maxima. This  tells us that randomness of networks plays some very interesting role. Completely regular networks are the best of all for reaching high degree of cooperation, completely random networks are only a little worse, and networks with a relatively low randomness of about 0.1 are the worst of all.

Fig. 9 indicates also that high density is always an obstacle for cooperation degree. Intuitively, this is not hard to understand: the more neighbors you have, the harder for you to cooperate.

Since the cellular automata network we study in the previous section is both completely regular and with a low density, no wonder that we observed an amazingly high degree of cooperation.

Now let's turn our attention from cooperation degree to another index, the average satisfaction degree. As discussed for the cellular automata network,
this index is always much smaller than cooperation degree. Calculation shows that its overall average value is 0.6, still significantly higher than the
initially expected value of 0.5, and confirms our claim again that best response dynamic promotes cooperation. As for the index of cooperation degree,
 high density of networks plays a negative role for average satisfaction. This effect is also quite robust. The effect of rewiring probability $q$,
 however, is unclear for low density networks ($k=8,18,28$). For high density networks ($k=38,48,58$), its effect is steadily negative. An interesting
 observation from the network with density $k=58$ is that, if we let $q$ grow from 0.1 to 1 and compare the index of cooperation degree and that of
 average satisfaction degree, we can find that the former index increases while the latter one decreases slowly. In this process, more and more agents
 are satisfied, but their average satisfaction degree becomes worse. This implies that more and more agents must have behaved compromisingly,
  leading the society to more equality. To put it another way, in the case that the network is highly dense and not so regular,
  adding randomness (or  decreasing regularity) of people's interactions can  improve  equality of the society.

It is valuable to remark that if we abandon the normalization in the definition of average satisfaction degree, i.e. define the {\it absolute} satisfaction degree of each agent as the number of neighbors that s/he likes, then we could find that adding density to people's interaction networks would be a good thing in general. This is not surprising, because having more neighbors means that you have more chances to get rewarded from interacting with them (notice that in the above argument the disadvantage from neighbors is not calculated).

As to the other two indices, i.e. equilibrium ratio and complete ratio, which are auxiliary in this paper, their overall average values are 0.45 and 0.09, respectively. Considering their definitions, these values are not low at all. The effects of rewiring probability $q$ and network density $k$, however, are not stable on neither of them.

\begin{figure}[!ht]
\begin{center}
\includegraphics[width=17cm]{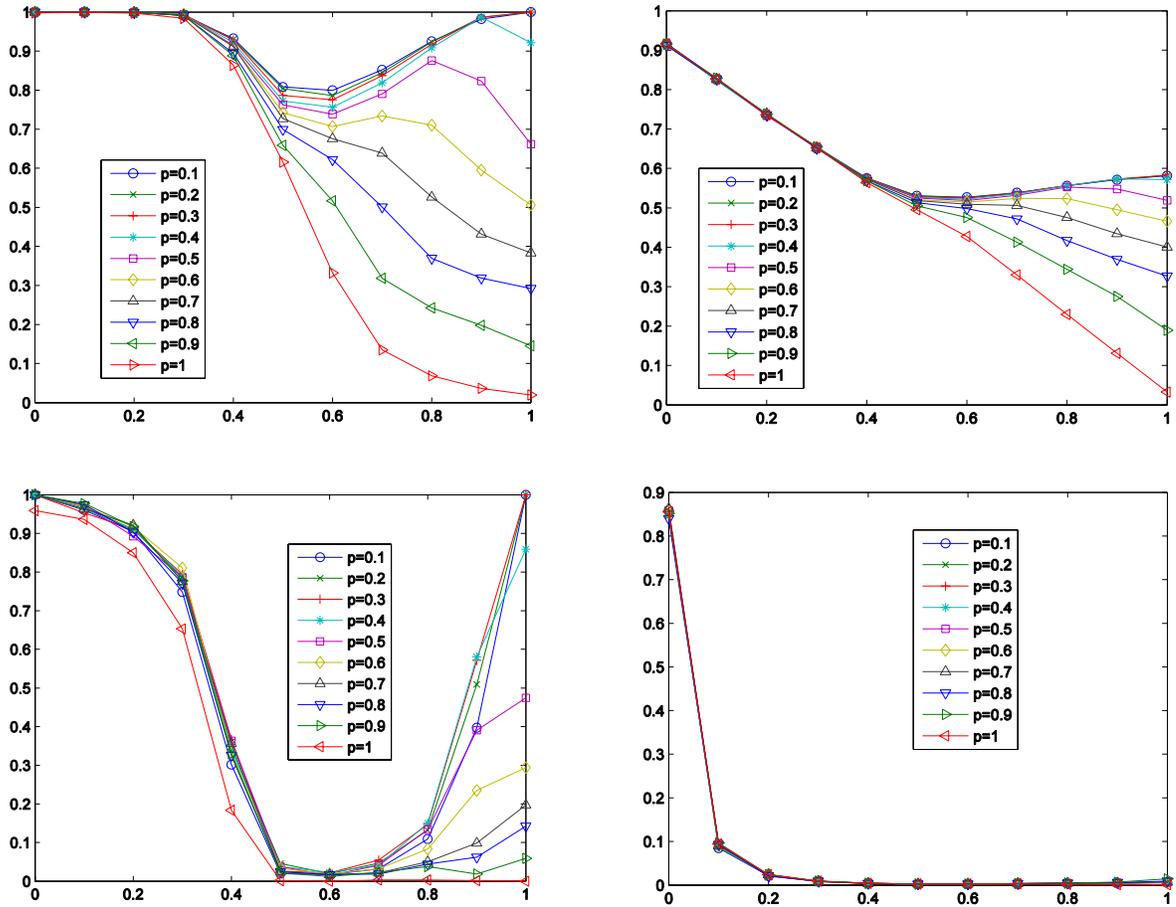}
\end{center}
\caption{
{\bf Effects of rebel ratio $r$ and updating probability $p$ on cooperation degree (Top-Left), average satisfaction degree (Top-Right), equilibrium ratio (Bottom-Left), and complete ratio (Bottom-Right).} All horizontal axises represent rebel ratio $r$.}
\label{Figure_10}
\end{figure}

Let's consider next the effects of rebel ratio $r$ and updating probability $p$. The simulation results, which excellently confirm our claims for the cellular automata network that high updating probability is always bad for cooperation and high rebel ratio is also bad when $p$ is large,  are displayed in Fig. 10. When $p$ is small, the effect of $r$ on cooperation level is roughly a pattern of ``V". This pattern can also be observed for the cellular automata network in Fig. 3. For small-world networks, however, it is much more significant, especially if we only concentrate on the index of equilibrium ratio. This result is plausible because in the two extreme cases with all conformists or all rebels, Nash equilibrium can always be guaranteed, but not in other cases. And the more balanced the proportion between conformists and rebels, the more difficult it is to reach a Nash equilibrium or a high average degree of satisfaction. It is quite remarkable that updating probability $p$ has absolutely no effect at all to the index of complete ratio. This ratio is very high for the case with all conformists. However, once the rebel ratio $r$ increases a little bit, it drops down dramatically, and then, very soon,  approximates zero.

Phase transition of equilibrium ratio is also confirmed in the more general settings, as displayed in Fig. 11. It should be noted that significance of this  phenomenon is negatively correlated with the rewiring probability $q$. For completely regular networks ($q=0$), phase transition is rather significant. However, as randomness of the network grows, this phenomenon becomes less and less significant. In fact, when $q$ is larger than 0.3, it is farfetched to call them phase transitions any more. Since the cellular automata network is completely regular, no wonder that we observed a beautiful phase transition.

\begin{figure}[!ht]
\begin{center}
\includegraphics[width=8.3cm]{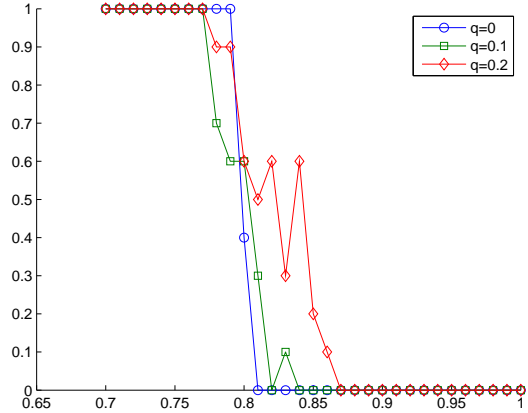}
\end{center}
\caption{
{\bf Phase transition of equilibrium ratio in updating probability $p$ in small-world networks ($k=8,r=1$).}The horizontal axis is for $p$.}
\label{Figure_11}
\end{figure}
\section*{Discussion}

The  fashion game is investigated in this paper through simulations. Our focus is on cooperation of agents. To study this, various indices are used, namely
the cooperation degree, the average satisfaction degree, the equilibrium ratio, and the complete ratio. Our finding is quite promising: in most cases,
agents can cooperate rather well through best response dynamics. It tells us also that the interaction structure matters a lot. Considering that these
dynamics are really simple, agents are selfish, myopic, naive, and have very limited information, this finding is fairly surprising.

It is valuable to note that small-world networks used in the last section are generated by the original algorithm of Watts and Strogatz. This algorithm,
though extremely popular and widely accepted, does not take the cellular automata network but a ring type regular network as the benchmark network ($p=0$).
As suggested by one anonymous referee, it is natural to take the cellular automata network as the benchmark network. This treatment will bridge the two
kinds of networks we study seamlessly. We call networks generated in this way the modified small-world networks. For these modified small-world networks
(with density $k=8$), we did the same group of simulations as in the previous section. For technical reasons, we didn't take the network size so large as
for the cellular automata network (recall that the size there is 1681), but a normal size of 441, about twice the size we used in the previous section for
the original small-world networks (200). We made a comparison between the original and the modified small-world networks. It turns out that, as shown in
Fig. 12, the two groups of networks make no meaningful difference for three of the four indices. This might indicate that, compared with network density
and randomness level, details such as how they are actually connected matter very little. Of course, this broader guess should be confirmed or falsified by
more extensive simulations, and related further studies include how degree distribution and clustering coefficient affect agents' behavior and consequently
their cooperation level in the fashion game. Network size, though does not affect cooperation degree, average satisfaction degree, or complete ratio
significantly, also as shown in Fig. 12, it may well matter a lot to equilibrium ratio. And the effect must be on the negative side. This is intuitive,
because the definition of an equilibrium is quite restrictive. As long as there is one agent that is unsatisfied, an action profile cannot be an
equilibrium. This possibility increases as the network size grows.

\begin{figure}[!ht]
\begin{center}
\includegraphics[width=8.3cm]{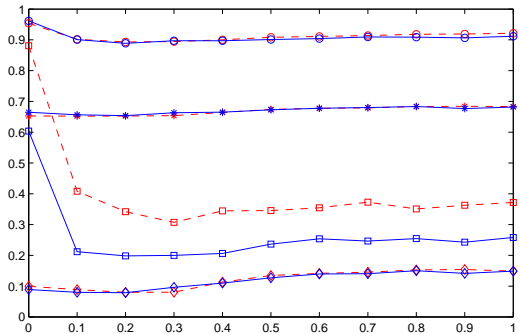}
\end{center}
\caption{{\bf Comparison between original small-world networks (dashed red lines) and modified small-world networks (solid blue lines).}The horizontal axis
is for rewiring probability $p$. $\circ$ is for the index of cooperation degree, $\ast$ for average satisfaction degree, $\square$ for equilibrium ratio,
and  $\lozenge$ for complete ratio.} \label{Figure_12}
\end{figure}

In this paper, we mainly focus on the overall cooperation levels, i.e. the macro side. For further research, it is also meaningful to investigate the micro
side of the fashion game, i.e. how the payoffs of agents are affected by their status in the networks, measured by various centrality indices, say their
degree centralities, their betweenness centralities, their closeness  centralities, and their clustering coefficients. To study these problems, the utility
function as defined in the introduction part of this paper should be given at least equal attention as the satisfaction degree.

Another obvious direction is to study the kind of realistic network, scale-free network \cite{ba99}. ``Homophily", a basic
 observation from the real world that ``birds of a feature flock together", is attracting more and more attention recently (c.f. \cite{msc01,ams09}), and also deserves
 the discussion of its effect on the fashion game. It is natural to conjecture that cooperation level is positively correlated with homophily
 level. Throughout this paper, we have assumed that the initial settings  are uniformly random, that is, each
agent takes action 1 with probability 0.5 and action 0 with probability 0.5. It is interesting to consider nonuniform initializations, say action 1 is
biased. Another interesting future direction is to take into account the possibility of network formation, where forming a new link (or severing an old
one, or both) is an extra strategy of each agent. Assuming that there is a cost to this new strategy, it is very meaningful to study the co-evolution of
cooperation level and network structure.

\section*{Acknowledgements} We thank professor Xiang Li
for reminding us of the connection between the anti-coordination game and the snowdrift game, and Dr. Yuqing Qiu for technical helps. We are also indebted
to two anonymous referees for their helpful suggestions and comments. In particular, more simulations on small-world networks and homophily and network
formation as interesting future directions were suggested by one of them, references \cite{w02} and \cite{lw08}  were also informed by her or him.
\bibliography{template}

\end{document}